\documentclass[12pt]{article}
\usepackage[margin=1.9cm]{geometry}
\usepackage{graphicx}
\usepackage{cite}
\usepackage{float}

\newcommand{\mysection}{\setcounter{equation}{0}\section}

\def\beq{\begin{equation}}
\def\eeq{\end{equation}}
\def\beqa{\begin{eqnarray}}
\def\eeqa{\end{eqnarray}}
 
\begin{document}

\begin{center}
{\Large \bf Higher-order QCD and electroweak corrections for $t{\bar t}H$ production}
\end{center}

\vspace{2mm}

\begin{center}

{\large Nikolaos Kidonakis$^a$ and Nodoka Yamanaka$^{b,c}$}\\

\vspace{2mm}

${}^a${\it Department of Physics, Kennesaw State University, \\
Kennesaw, GA 30144, USA}

\vspace{1mm}

${}^b${\it Department of Physics, Tohoku University, \\ 
Sendai 980-8578, Japan}

\vspace{1mm}

${}^c${\it Nishina Center for Accelerator-Based Science, RIKEN, \\
Wako 351-0198, Japan}

\end{center}

\begin{abstract}
We calculate higher-order soft-gluon corrections for the associated production of a top-antitop quark pair and a Higgs boson ($t{\bar t}H$ production) using resummation in one-particle-inclusive (1PI) kinematics. By adding these corrections to NLO QCD and NLO electroweak (EW) results, we present cross sections through approximate N$^3$LO (aN$^3$LO) in QCD and NLO in EW for $t{\bar t}H$ production at LHC energies. These aN$^3$LO QCD + NLO EW results provide significant enhancements over lower orders and a reduced scale dependence. Based on our results and current experimental data, we derive a constraint on the deviation of the top quark Yukawa coupling from the Standard Model value which is a probe of new physics beyond it. We also calculate top-quark differential distributions in transverse momentum and rapidity.
\end{abstract}

\mysection{Introduction}

The production of a top-antitop quark pair in association with a Higgs boson is a process that has been studied intensely theoretically and experimentally for a long time, and it has been observed by both the ATLAS and the CMS experiments at the LHC \cite{ATLASe,CMSe,CMS1,ATLAS1,CMScp,ATLAS2,CMS2,ATLAS3}. The process involves the two heaviest elementary particles in the Standard Model and it provides the possibility of a measurement of the Yukawa coupling of the top quark. Determining the Yukawa interactions is not only important as a confirmation of the Standard Model, but it may also probe candidates of new physics beyond it, such as the two-Higgs-doublet models \cite{GCB}, supersymmetric models \cite{HHGK,JGHH}, composite models \cite{DGK,ACP}, the Froggatt-Nielsen model \cite{CFHN}, etc. 

The precision of the theoretical predictions for this process has increased a lot over time. The leading-order (LO) QCD prediction for $t{\bar t}H$ production was presented in Refs. \cite{LO1,LO2}. The next-to-leading-order (NLO) QCD corrections to the $t{\bar t}H$ cross section were calculated in Refs. \cite{NLO1,NLO2,NLO3,NLO4,NLO5,NLO6} (see also \cite{CLE}, and NLO QCD corrections for $t{\bar t} Hj$ in Ref. \cite{ttHj}). The NLO electroweak (EW) corrections for $t {\bar t} H$ production were calculated in Refs. \cite{EW1,EW2,EW3}. Furthermore, soft-gluon corrections from threshold resummation were presented using various approaches in Refs. \cite{res1,res2,res3,res4,res5,res6,res7}.

Calculations with top-quark and Higgs decays and parton showers appeared in Refs. \cite{dps1,dps2,dps3,dps4}, with decays at fixed-order within the narrow width approximation (NWA) in Ref. \cite{EW2}, with non-resonant contributions and off-shell effects at NLO QCD in Ref. \cite{rpnf1} and with NLO EW corrections in Ref. \cite{rpnf2}. NLO QCD predictions for full off-shell $t{\bar t}H$ production with Higgs boson decay via the NWA at NLO QCD appeared in Refs. \cite{os1,os2}.

Progress towards $t{\bar t}H$ production to next-to-next-to-leading-order (NNLO) QCD accuracy has been made recently in Refs. \cite{NNLO1,NNLO2,NNLO3,CFKBR,WXYY,AHJKK,NNLO4,NNLO5}. The latest result in \cite{NNLO5} combines partial NNLO QCD predictions with EW corrections and soft-gluon resummation. The resummation contributions in \cite{NNLO5} are presented with two distinct resummation formalisms, but the common approach in these two formalisms as applied to this process is the choice of kinematics, i.e. the invariant mass of the final state. Here, in contrast, we present a calculation using one-particle-inclusive (1PI) kinematics.

The resummation formalism that we employ here has been developed and used for many other top-quark processes, including $t{\bar t}$ \cite{NKGS1,NKGS2,NKtt2l} and single top \cite{NKnlltev,NKnlllhc,NKsch,NKtW,NKtch,NK3l,NKs25} production, as well as top-quark processes with three-particle final states such as $tqH$ \cite{MFNK,tqH}, $tq\gamma$ \cite{tqgamma}, $tqZ$ \cite{tqZ}, $t{\bar t}\gamma$ \cite{ttgamma}, $t{\bar t}W$ \cite{ttW}, and $t{\bar t}Z$ \cite{ttZ} production.

In the next section, a brief overview of the higher-order corrections, including the soft-gluon resummation formalism, is given. In Section 3, we provide numerical results for $t{\bar t}H$ total cross sections at LHC energies. In Section 4, we present top-quark transverse-momentum ($p_T$) and rapidity distributions. A comparison with experimental data is provided in Section 5. We conclude in Section 6.

\mysection{Higher-order corrections for $t{\bar t}H$ production}

At leading order, the partonic processes for $t{\bar t}H$ production are $a(p_a)+b(p_b) \to t(p_t)+{\bar t}(p_{\bar t})+H(p_H)$, where $a$ and $b$ denote quarks and antiquarks or gluons. We define the usual kinematical variables $s=(p_a+p_b)^2$, $t=(p_a-p_t)^2$, and $u=(p_b-p_t)^2$. The perturbative QCD corrections beyond LO to $t{\bar t}H$ production are significant. As already discussed in the introduction, complete NLO QCD \cite{NLO1,NLO2,NLO3,NLO4,NLO5,NLO6} and NLO EW \cite{EW1,EW2,EW3} corrections have been available for a while, and significant progress has been made more recently at NNLO QCD \cite{NNLO1,NNLO2,NNLO3,CFKBR,WXYY,AHJKK,NNLO4,NNLO5}. 

An important subset of the QCD corrections arise from soft-gluon emission near partonic threshold, where there is little energy for additional radiation. We will use 1PI kinematics, with the top quark being the observed particle with mass $m_t$. Considering the emission of a gluon with momentum $p_g$, we define the partonic threshold variable $s_4=(p_{\bar t}+p_H+p_g)^2-(p_{\bar t}+p_H)^2=s+t+u-m_t^2-(p_{\bar t}+p_H)^2$, which vanishes when $p_g \to 0$. At each order in the perturbative series, soft-gluon contributions appear in the analytical expressions as terms which consist of coefficients multiplying logarithms of the form $\ln^k(s_4/m_t^2)/s_4$, where the integer $k$ takes values from 0 through $2n-1$ in the $n$th-order corrections.

After Laplace transforms, one expresses the factorized form of the 1PI cross section as a product of a hard-scattering term and functions that describe collinear and soft gluon emission. The renormalization-group evolution of these functions \cite{NKGS1,NKGS2,NKtt2l,NKnlltev,NKnlllhc,NKsch,NKtW,NKtch,NK3l,NKs25,MFNK,tqH,tqgamma,tqZ,ttgamma,ttW,ttZ,GS,LOS,NK2loop,NK4loop} results in resummation, i.e. exponentiation, of these corrections. Details are given in the related paper on $t{\bar t}W$ production \cite{ttW} (see also \cite{ttZ} on $t{\bar t}Z$ production) for which the resummation formalism is the same as in the present case. We expand the resummed cross section to fixed order, and then invert back to momentum space to derive fixed-order results without need for a prescription \cite{NKtt2l,NKnlltev,NKnlllhc,NKsch,NKtW,NKtch,NK3l,NKs25,MFNK,tqH,tqgamma,tqZ,ttgamma,ttW,ttZ}. By adding second-order soft-gluon corrections to the exact NLO QCD result, we obtain approximate NNLO (aNNLO) QCD predictions. By further adding third-order soft-gluon corrections, we obtain approximate N$^3$LO (aN$^3$LO) QCD predictions. We note that with a next-to-next-to-leading-logarithm (NNLL) resummation, the terms in the last two logarithmic powers of the aN$^3$LO corrections cannot be fully determined.

As discussed in some detail in \cite{ttW} and \cite{ttZ}, in comparing different approaches to resummation, it is important not only to note the broad framework of the formalism (e.g. resummation in QCD under Laplace transforms vs resummation in soft-collinear effective theory) but also to specify the kinematics and the variable that is used in the logarithmic terms. In the 1PI kinematics that is employed in this paper, we have logarithms of $s_4/m_t^2$ in the perturbative series, while previous works have used a variable involving absolute threshold \cite{res1,res5} or the invariant mass of the $t{\bar t}H$ system \cite{res2,res3,res4,res6,res7,NNLO5}. Use of different variables can have a significant numerical impact; furthermore, the choice of kinematics guides the choice of the central factorization and renormalization scales, which we take as the mass of the observed particle, i.e. the top quark, in 1PI kinematics. 

A solid test of the relevance of a resummation formalism for processes where soft-gluon corrections are numerically dominant (such as top-quark processes) is its prediction of and agreement with exact fixed-order calculations at higher orders. The formalism used in this study was highly successful \cite{NKtt2l} in predicting the $t{\bar t}$ cross section at NNLO to very high accuracy. More recently, in Ref. \cite{ttW}, the same was shown for $t{\bar t}W$ production; we find in the present study that the same holds for $t{\bar t}H$ production.

\mysection{$t{\bar t}H$ cross sections}

In this section, we provide numerical results for the total cross sections for $t{\bar t}H$ production at LHC energies. In our results we use $m_t=172.5$ GeV, $m_H=125.2$ GeV \cite{RPP24}, and MSHT20 aN$^3$LO parton distribution functions (pdf) \cite{MSHT20an3lo}. We set the central factorization and renormalization scales equal to the top-quark mass. 

\begin{table}[htbp]
\begin{center}
\begin{tabular}{|c|c|c|c|c|c|c|} \hline
\multicolumn{4}{|c|}{$t{\bar t}H$ cross sections in $pp$ collisions at the LHC} \\ \hline
$\sigma$ in fb & 13 TeV & 13.6 TeV & 14 TeV \\ \hline
LO QCD & $434^{+138}_{-97}{}^{+9}_{-9}$ & $485^{+152}_{-108}{}^{+10}_{-10}$ & $520^{+162}_{-115}{}^{+10}_{-11}$ \\ \hline
NLO QCD & $497^{+14}_{-38}{}^{+11}_{-12}$ & $558^{+17}_{-43}{}^{+12}_{-13}$ & $601^{+18}_{-46}{}^{+13}_{-14}$ \\ \hline
NLO QCD + NLO EW & $503^{+15}_{-37}{}^{+11}_{-12}$ & $564^{+17}_{-42}{}^{+13}_{-13}$ & $607^{+19}_{-45}{}^{+13}_{-14}$ \\ \hline
aNNLO QCD & $518^{+4}_{-14}{}^{+12}_{-13}$ & $581^{+4}_{-15}{}^{+13}_{-14}$ & $626^{+5}_{-16}{}^{+14}_{-15}$ \\ \hline
aNNLO QCD + NLO EW & $524^{+5}_{-13}{}^{+12}_{-13}$ & $587^{+5}_{-14}{}^{+13}_{-14}$ & $632^{+6}_{-15}{}^{+14}_{-15}$ \\ \hline
aN$^3$LO QCD & $523^{+3}_{-7}{}^{+12}_{-13}$ & $587^{+3}_{-7}{}^{+13}_{-14}$ & $633^{+3}_{-8}{}^{+14}_{-15}$ \\ \hline
aN$^3$LO QCD + NLO EW & $529^{+3}_{-6}{}^{+12}_{-13}$ & $593^{+3}_{-6}{}^{+13}_{-14}$ & $639^{+3}_{-7}{}^{+14}_{-15}$ \\ \hline
\end{tabular}
\caption[]{The $t{\bar t}H$ cross sections (in fb) with QCD and EW corrections at various orders, with scale and pdf uncertainties, in $pp$ collisions with $\sqrt{S}=13$, 13.6, and 14 TeV, $m_t=172.5$ GeV, $m_H=125.2$ GeV, and MSHT20 aN$^3$LO pdf.}
\label{table1}
\end{center}
\end{table}

In Table 1, we present the $t{\bar t}H$ cross sections at various QCD and EW orders at 13, 13.6, and 14 TeV LHC energies. We show results at LO QCD, NLO QCD, NLO QCD + NLO EW, aNNLO QCD, aNNLO QCD + NLO EW, aN$^3$LO QCD, and aN$^3$LO QCD + NLO EW. The cross sections through NLO QCD + NLO EW are calculated using {\small \sc MadGraph5\_aMC@NLO} \cite{MG5,MGew}. To those results we add second-order and third-order soft-gluon corrections to derive the higher-order cross sections. We see that all perturbative corrections are positive. The aN$^3$LO QCD + NLO EW result is 22\% higher than LO QCD.

The first set of uncertainties for each cross section is from scale variation over the interval $m_t/2 \le \mu \le 2m_t$. The uncertainties are very similar for the three energies given in the table. We note that it makes no difference whether we vary the factorization and renormalization scales simultaneously (3-point variation) or if we use 7-point variation, as also reported for the related processes of $t{\bar t}$ production in Ref. \cite{KGT} and for $t{\bar t}W$ production in Ref. \cite{ttW}. While the scale variation is huge at LO (+31\% -22\%), it becomes much smaller at NLO, and it gets further reduced at higher orders (down to around one percent). The second set of uncertainties is from the parton distributions. The pdf uncertainties are also very similar for the three energies and, furthermore, they do not depend much on the perturbative order (they are roughly $\pm$ 2\% for all energies and orders). Thus, the pdf uncertainties are much smaller than the scale uncertainties at LO, and still smaller at NLO, but they are bigger than the scale uncertainties from aNNLO through aN$^3$LO QCD + NLO EW.

Though, as discussed earlier, we used different kinematics and scale choices than other works on the subject, we find that our aNNLO results are consistent, within theoretical uncertainties, with the partial NNLO results in \cite{NNLO3,NNLO4,NNLO5} (see also the discussion in Section 5). We note that the central scale used for the total cross section in Refs. \cite{NNLO3,NNLO4,NNLO5} is $m_t+m_H/2$. The variation by factor of 2 that is employed means that the scale variation in those works is between $m_t/2+m_H/4$ and $2m_t+m_H$. In contrast, the central scale in our calculation is $m_t$ and, thus, the scale variation is between $m_t/2$ and $2m_t$, which is a smaller span than that in \cite{NNLO3,NNLO4,NNLO5}. Therefore, the uncertainty due to scale variation is somewhat smaller in our result (as one would expect). Furthermore, if we use the same pdf and other parameters as in \cite{NNLO3,NNLO4,NNLO5}, we find that our central aNNLO results are slightly higher than those in \cite{NNLO3,NNLO4,NNLO5}, which is again consistent with the fact that our central scale is smaller.

\mysection{Top-quark $p_T$ and rapidity distributions}

In this section, we provide numerical results for the top-quark $p_T$ and rapidity distributions in $t{\bar t}H$ production at 13 TeV and 13.6 TeV LHC energies. As before, in our results we use $m_t=172.5$ GeV, $m_H=125.2$ GeV, and MSHT20 aN$^3$LO pdf \cite{MSHT20an3lo}.

\begin{figure}[htbp]
\begin{center}
\includegraphics[width=88mm]{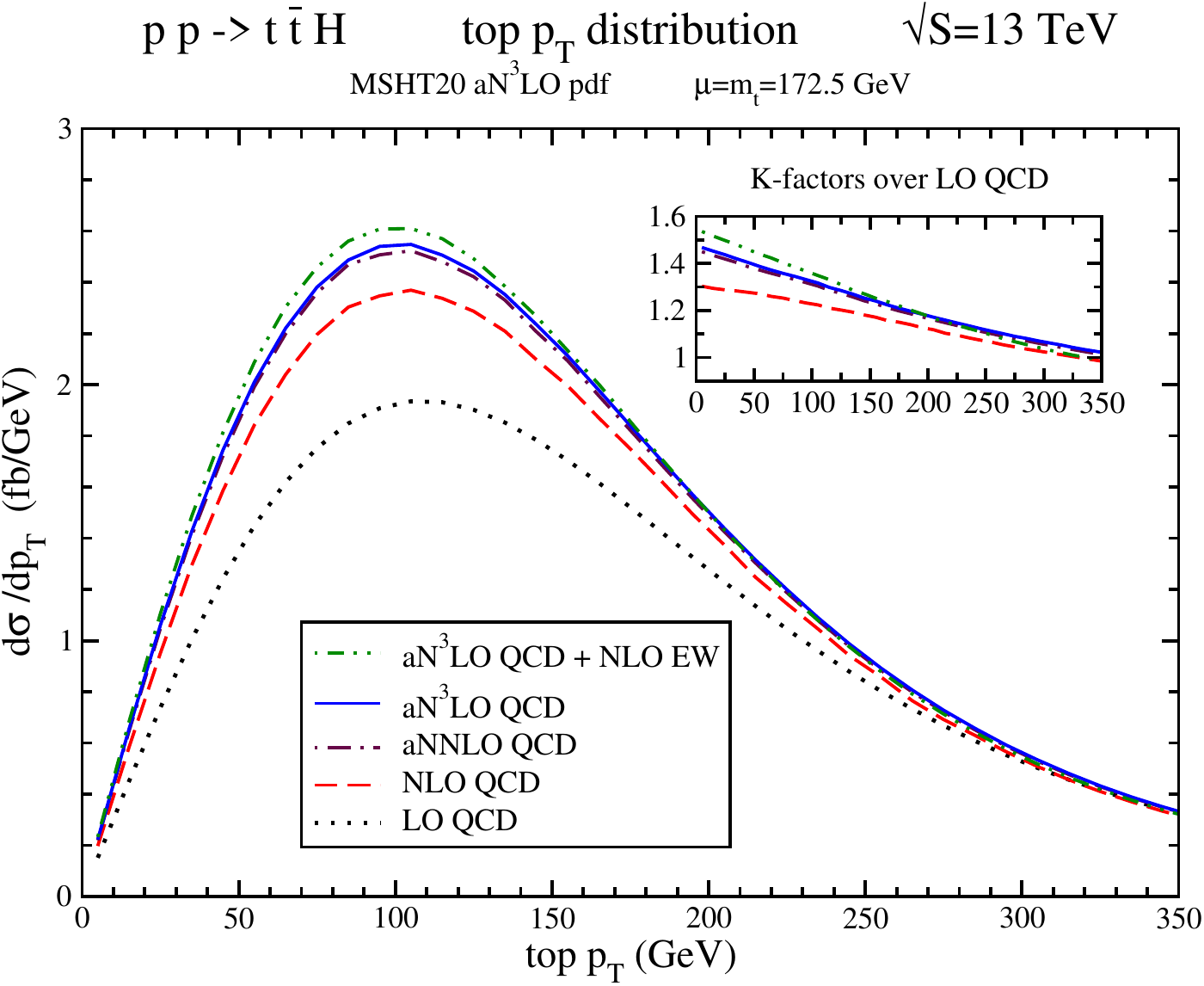}
\includegraphics[width=88mm]{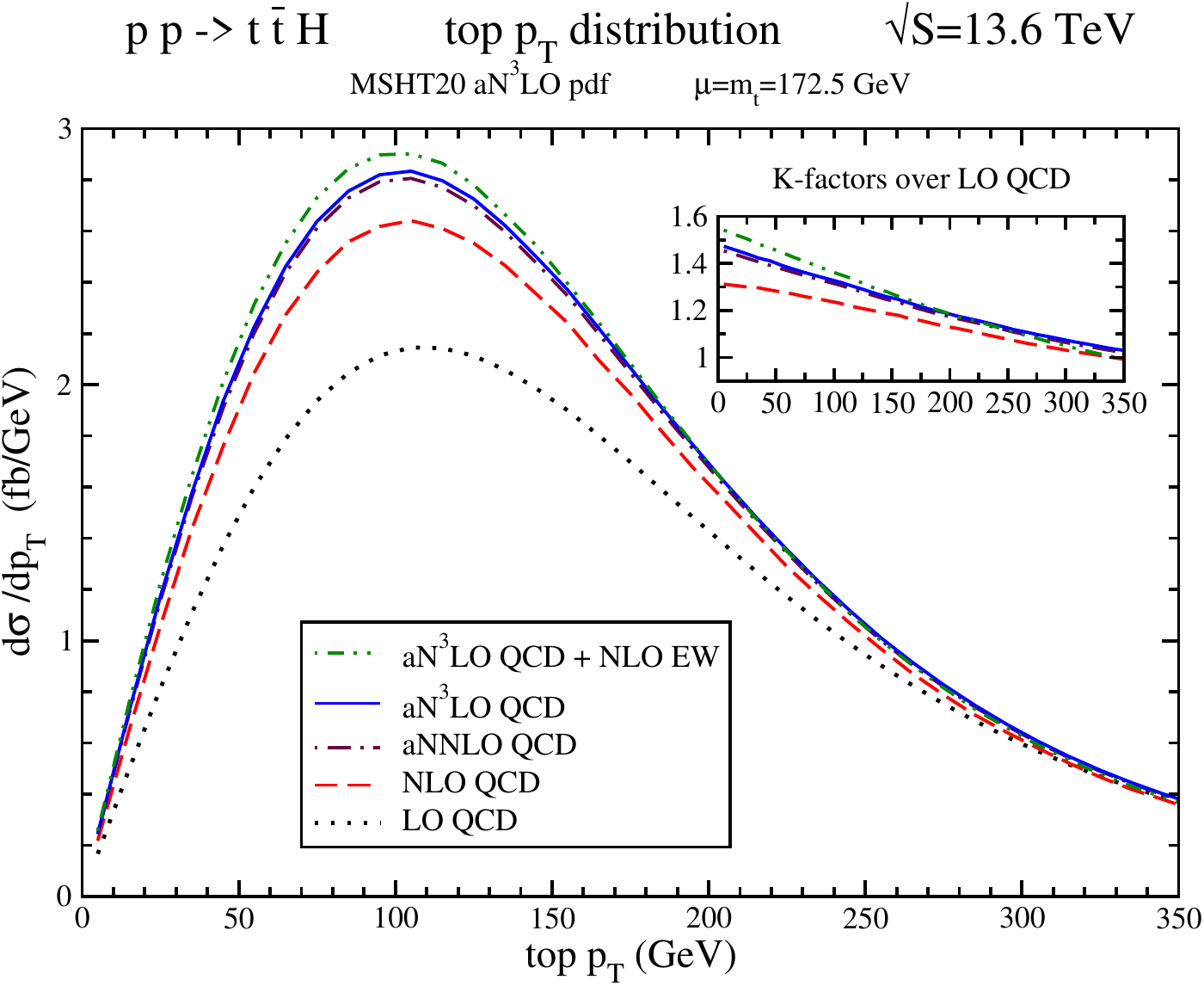}
\caption{The top-quark $p_T$ distributions through aN$^3$LO QCD + NLO EW in $t{\bar t}H$ production in $pp$ collisions at an LHC energy of 13 TeV (left plot) and 13.6 TeV (right plot). The inset plots display the $K$-factors relative to LO.}
\label{pttopplot}
\end{center}
\end{figure}

In Fig. \ref{pttopplot}, we show the top-quark $p_T$ distribution in $t{\bar t}H$ production at 13 TeV and 13.6 TeV LHC energies, at various orders from LO QCD through aN$^3$LO QCD + NLO EW, with MSHT20 aN$^3$LO pdf. The distributions peak at $p_T$ values of slightly above 100 GeV at all orders for both energies. The inset plots show the $K$-factors, i.e. the ratios of the higher-order results to LO QCD. The $K$-factors are large at smaller $p_T$ and gradually decrease at larger $p_T$ values. As can be seen from the plots, the $K$-factors at 13.6 TeV are very similar to those at 13 TeV. Finally, we note that the scale uncertainties of the  aN$^3$LO QCD + NLO EW $p_T$ distributions are similar to those for the total cross section.

\begin{figure}[htbp]
\begin{center}
\includegraphics[width=88mm]{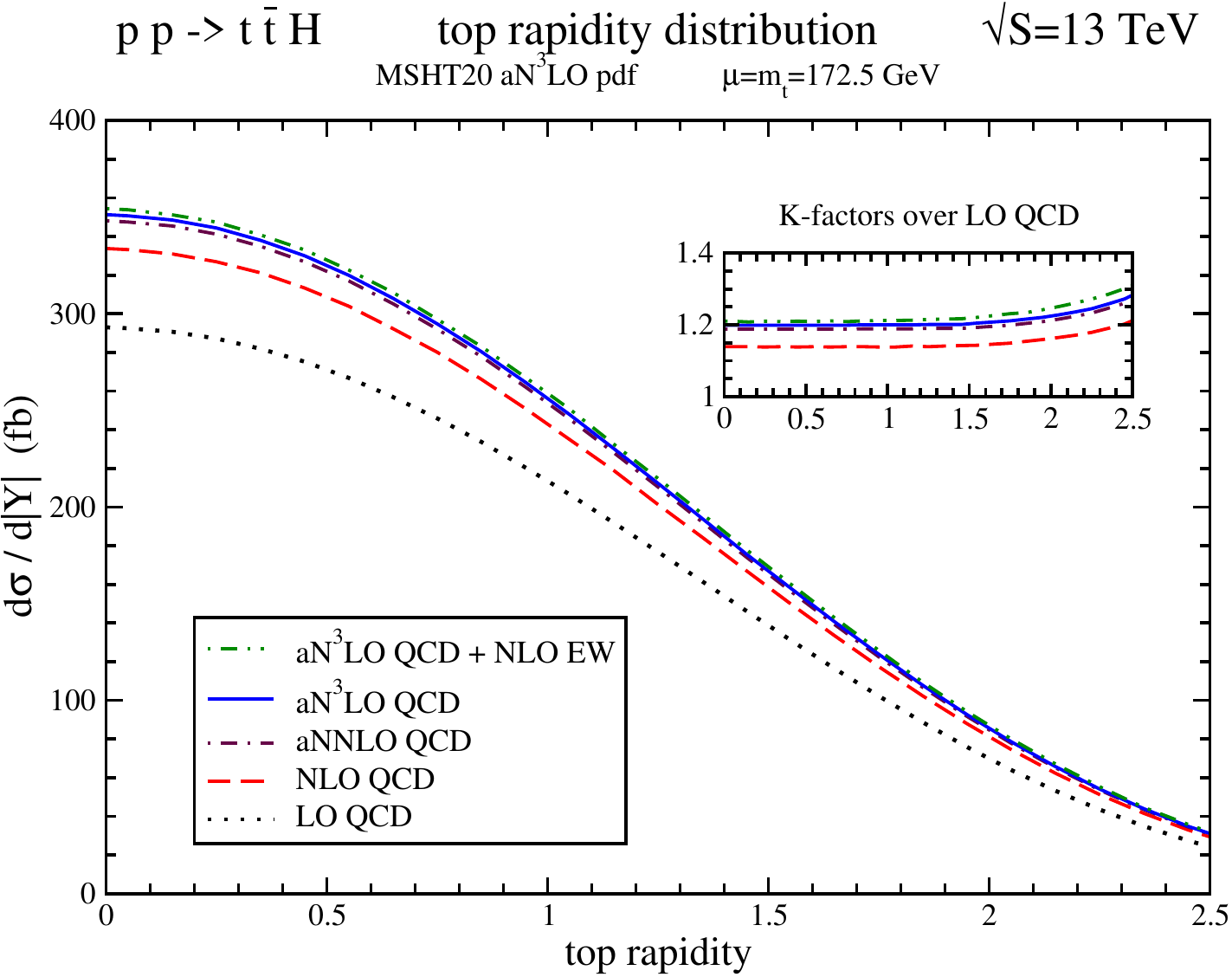}
\includegraphics[width=88mm]{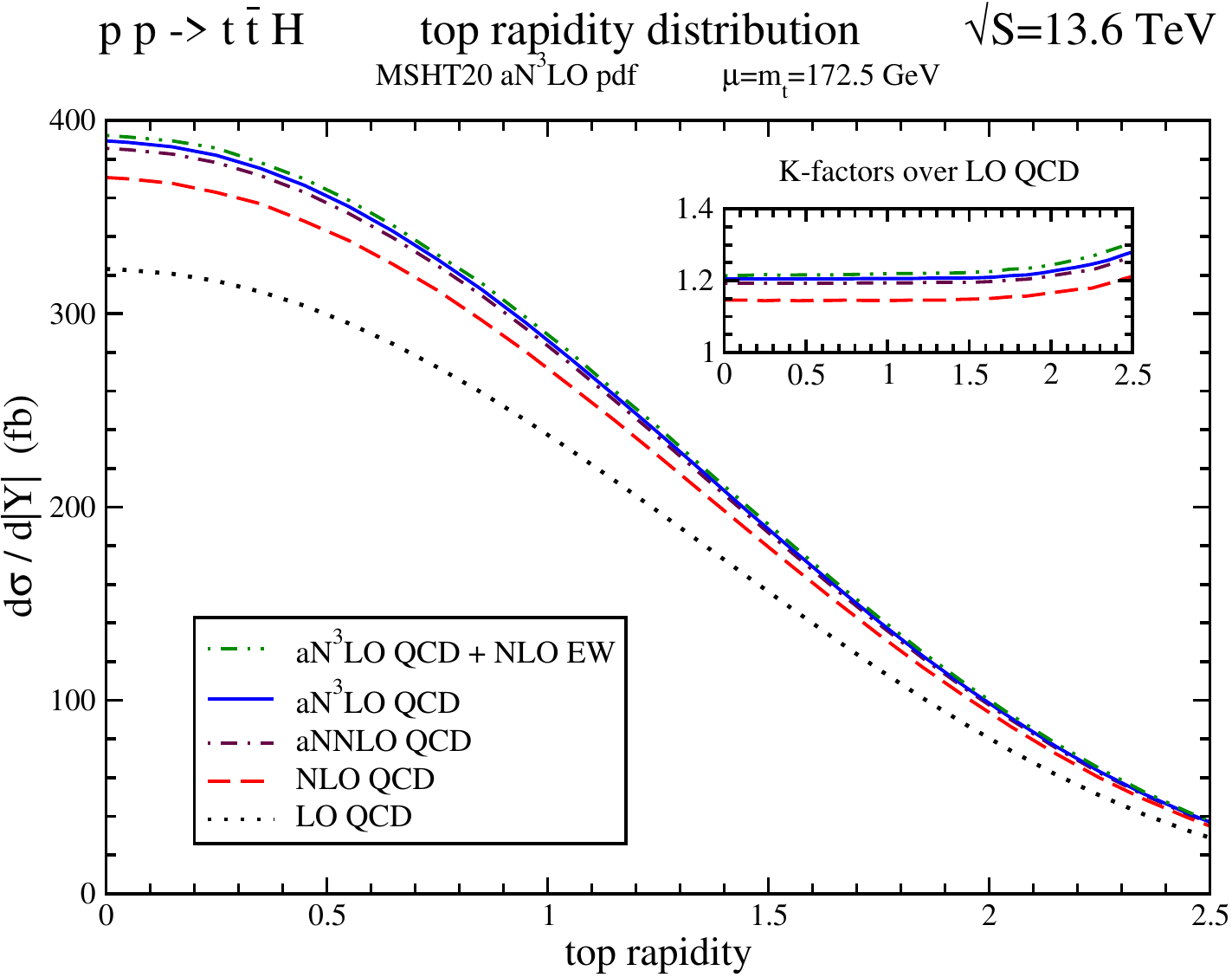}
\caption{The top-quark rapidity distributions through aN$^3$LO QCD + NLO EW in $t{\bar t}H$ production in $pp$ collisions at an LHC energy of 13 TeV (left plot) and 13.6 TeV (right plot). The inset plots display the $K$-factors relative to LO.}
\label{ytopplot}
\end{center}
\end{figure}

In Fig. \ref{ytopplot}, we show the top-quark rapidity distribution in $t{\bar t}H$ production at 13 TeV and 13.6 TeV LHC energies, again at various orders from LO QCD through aN$^3$LO QCD + NLO EW, using MSHT20 aN$^3$LO pdf. The inset plots show the $K$-factors relative to LO QCD. As can be seen from the plots, the $K$-factors increase at larger rapidities, and they are very similar at the two energies. Finally, we note that the scale uncertainties of the aN$^3$LO QCD + NLO EW rapidity distributions are similar to those for the total cross section.

\mysection{Analysis and comparison to data}

We now analyze the results of our calculations and compare our output with the data from the LHC experiments. We begin with the results from ATLAS. In their paper \cite{ATLASe} on evidence for $t{\bar t}H$ production, ATLAS measured a cross section of $790^{+230}_{-210}$ fb at 13 TeV assuming a Higgs mass of 125 GeV. In their later paper \cite{ATLAS1} on the observation of $t{\bar t}H$ production, ATLAS measured a cross section of $670 \pm 90 {}^{+110}_{-100}$ fb at 13 TeV using a Higgs mass of 125.09 GeV. The latest cross section measurement from ATLAS at 13 TeV \cite{ATLAS3} is $411^{+101}_{-92}$ fb assuming a Higgs mass of 125.09 GeV. The results have large error bars but are consistent with the theoretical predictions.

Next, we discuss the results from CMS. In their paper \cite{CMSe} on evidence for $t{\bar t}H$ production at 13 TeV, CMS measured a signal rate of $1.23^{+0.45}_{-0.43}$ times the production rate expected in the Standard Model for a Higgs mass of 125 GeV. In their paper \cite{CMS1} on the observation of $t{\bar t}H$ production based on a combined analysis of data at 7, 8, and 13 TeV, CMS measured a signal rate of $1.26^{+0.31}_{-0.26}$ times the production rate expected in the Standard Model for a Higgs mass of 125.09 GeV. In their latest paper \cite{CMS2} on $t{\bar t}H$ production at 13 TeV, CMS has calculated a best fit value of $0.33 \pm 0.26$ relative to a theoretical expectation of $507^{+35}_{-50}$ fb assuming a Higgs mass of 125 GeV. Again, the uncertainties in the experimental results to date have been very large.

\begin{figure}[htbp]
\begin{center}
\includegraphics[width=100mm]{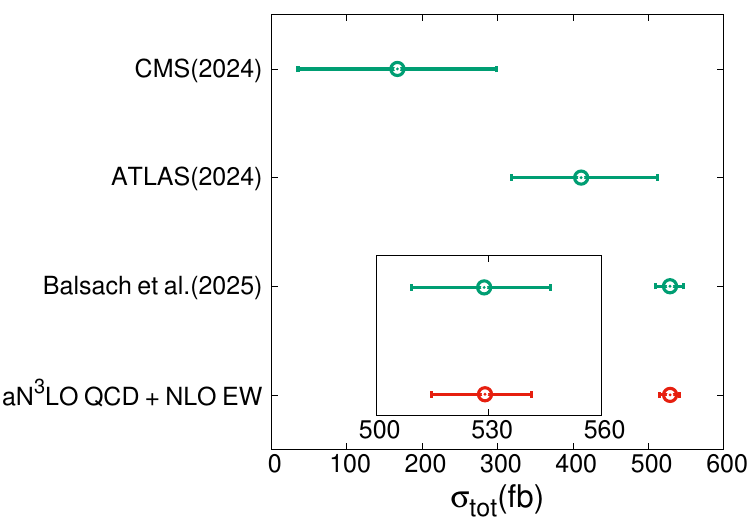}
\caption{Comparison of the total cross section of $t{\bar t}H$ production in 13 TeV $pp$ collision between the experimental data of CMS \cite{CMS2}, ATLAS \cite{ATLAS3}, and our aN$^3$LO QCD + NLO EW result.
The theoretical result of Balsach {\it et al.} \cite{NNLO5} is also shown. The inset plot affords a clearer comparison of our result with that of Ref. \cite{NNLO5}.}
\label{comparisonplot}
\end{center}
\end{figure}

In Fig. \ref{comparisonplot}, we plot our aN$^3$LO QCD + NLO EW results together with the experimental data of CMS \cite{CMS2} and ATLAS \cite{ATLAS3}. We see that the ATLAS result is marginally consistent with ours, while the CMS one is affected by a large uncertainty. We also compare to the result from Ref. \cite{NNLO5} and find agreement within the theoretical errors.

Next, we try to derive constraints on new physics beyond the Standard Model.
By assuming that our calculation is proportional to the square of the top quark Yukawa coupling, $Y_t^2$, it is possible to derive an upper limit to its deviation $\delta Y_t$ from the Standard Model value.
A naive comparison with the ATLAS data \cite{ATLAS3} yields
\beq
| \delta Y_t | \le 0.44 \, Y_t \, .
\eeq
We note that $Y_t$ is much more tightly constrained by the gluon-fusion Higgs production ($gg \to H$) cross section, also measured at the LHC, because the top quark loop represents its dominant contribution.
The consistency of the gluon fusion between the Standard Model and the experimental results \cite{ATLAS4} therefore means that any deviations of $Y_t$ from the Standard Model imply the existence of an additional contribution to the $ggH$ vertex beyond the Standard Model to counterbalance the deficit.
We may then state that the deviation of the $t {\bar t} H $ production cross section from the Standard Model value is a probe not only of the shift of the top quark Yukawa coupling $\delta Y_t$, but also to the new $ggH$ vertex.
This effective $ggH$ interaction may be generated by a loop, or a sum of loops of colored particles beyond the Standard Model coupled to the Higgs boson, which is not an exceptionally ad hoc feature.
The degree of the fine tuning of the coupling constants is of the order of $O( |\delta Y_t / Y_t|)$, so further improvements of the experimental accuracy will validate or reject the contribution of the new physics from the perspective of the naturalness.
Another option is to compare our $p_T$ or rapidity differential cross sections with the experimental data to extract the deviation of $Y_t$, but there are currently no experimental data available.
This is an interesting point to be considered in the future.

\mysection{Conclusions}

We have presented theoretical predictions for $t{\bar t}H$ cross sections at LHC energies. We have included complete QCD and EW corrections at NLO and, in addition, approximate NNLO and N$^3$LO corrections from soft-gluon resummation in 1PI kinematics. The aN$^3$LO QCD + NLO EW results provide significant enhancements to the lower-order cross sections with a large reduction in theoretical uncertainty from scale dependence. 

We have compared our results to data from the ATLAS and CMS experiments at the LHC. The current experimental uncertainties are very large. We have used the comparison to the data to derive constraints on new physics beyond the Standard Model.

We have also presented top-quark differential distributions in transverse momentum and rapidity. Again, the higher-order corrections are significant and they depend on the $p_T$ and rapidity values of the top quark.

\section*{Acknowledgements}
The work of N.K. is supported by the National Science Foundation under Grant No. PHY 2412071. The work of N.Y. is supported by the RIKEN TRIP initiative (Nuclear transmutation), and also partially by ERATO project-JPMJER2304.

\end{document}